\documentstyle[preprint,prl,aps,epsf]{revtex}
\begin{document}
\title{Shadowing Effects on the Nuclear Suppression Factor, $R_{\rm dAu}$, in 
d+Au Interactions}

\author{R. Vogt}
 
\address{{
Nuclear Science Division, Lawrence Berkeley National Laboratory, 
Berkeley, CA 94720, USA\break 
and\break
Physics Department, University of California, Davis, CA 95616, USA}\break}
 
\vskip .25 in
\maketitle
\begin{abstract}

We explore how nuclear modifications to the nucleon parton
distributions affect production of high transverse momentum 
hadrons in deuteron-nucleus
collisions.  We calculate the charged hadron spectra to leading order
using standard fragmentation
functions and shadowing parameterizations.  We obtain the d+Au to $pp$ ratio
both in minimum bias collisions and as a function of centrality.  The minimum
bias results agree reasonably 
well with the BRAHMS data while the calculated centrality dependence 
underestimates the data and is a 
stronger function of $p_T$ than the data indicate.
\end{abstract}

One of the most intriguing results from the
Relativistic Heavy Ion Collider (RHIC) at Brookhaven National
Laboratory has been the suppression of hadrons with high transverse momentum, 
$p_T$, in central Au+Au collisions at center-of-mass energies,
$\sqrt{S_{NN}}$,  of 130 and 200 GeV.  The $AA$ suppression factor,
\begin{equation}
R_{AA}(p_T)  = {d \sigma_{AA}/dp_T \over \langle \sigma_{NN}^{\rm in} T_{AA} 
\rangle d\sigma_{pp}/dp_T } \, \, , 
\label{raadef}
\end{equation}
compares the $AA$ and $pp$ $p_T$ distributions of hadrons, 
normalized by the number
of binary collisions, $\langle \sigma_{NN}^{\rm in} T_{AA} 
\rangle$, the product of the nucleon-nucleon
inelastic cross section, $\sigma_{NN}$,
and the nuclear overlap function $T_{AA}$.  Saturation effects in the initial
nuclear wave function and final-state parton energy loss were both proposed as 
explanations of the
large suppression seen in Au+Au collisions by PHENIX\cite{phenix130,phenix200},
STAR \cite{star130,star200} 
and BRAHMS \cite{brahms}.  To determine whether the suppression
is an initial or final-state effect, d+Au collisions at $\sqrt{S_{NN}} = 200$
GeV were recently studied at RHIC.  The data \cite{phenixdau,stardau,brahmsdau}
show that, at midrapidity $(\eta \approx 0)$, the d+Au suppression factor,
\begin{equation}
R_{\rm dAu}(p_T)  = {d \sigma_{\rm dAu}/dp_T \over
\langle \sigma_{NN}^{\rm in} T_{\rm dAu} \rangle d\sigma_{pp}/dp_T } \, \, ,
\label{rdadef}
\end{equation}
is much closer to unity.  These results
suggest that the strong suppression in Au+Au collisions is a final-state
effect, implying that, at least at central rapidities, saturation effects are
small.  However, at higher rapidities where the nuclear parton momentum
fraction, $x_2$, is smaller, such effects might still be important.  Since $x$
is not very small at RHIC, it is necessary to check if other, more
conventional, models of nuclear shadowing may also explain the data.  

The BRAHMS collaboration has measured $R_{\rm dAu}$ at several values of
pseudorapidity, $\eta$, and observed increasing suppression as $\eta$ increases
from $|\eta|\leq 0.2$ to $\eta = 3.2$ \cite{brahmsdau}.
The BRAHMS measurements are in four $\eta$ bins: $|\eta|\leq 0.2$; $0.8 \leq
\eta \leq 1.2$ ($\eta = 1$); $1.9 \leq \eta \leq 2.35$ ($\eta = 2$) and $2.9
\leq \eta \leq 3.5$ ($\eta = 3.2$), corresponding to center-of-mass scattering
angles, $\theta_{\rm cm}$, of $101.4^\circ \geq 
\theta_{\rm cm} \geq 78.6^\circ$, $48.4^\circ \geq \theta_{\rm cm} \geq
33.5^\circ$, $17.01^\circ \geq \theta_{\rm cm} \geq 10.9^\circ$ and $6.3^\circ
\geq \theta_{\rm cm} \geq 3.5^\circ$ respectively.
These data have also been divided into three centrality bins:
$(0-20)$\%, $(30-50)$\% and $(60-80)$\% of the geometric cross section.
Using a Glauber calculation of the nuclear overlap with a Woods-Saxon density
distribution for the gold nucleus and the Hulth\'en wavefunction to calculate
the deuteron density, we find that these bins correspond to the impact 
parameter, $b$, ranges of $0 \leq b \leq 3.81$ fm, $4.66 \leq b \leq 6.01$ fm
and $6.59 \leq b \leq 7.74$ fm respectively.  Our calculated values of 
$\langle \sigma^{\rm in}_{NN} T_{\rm dAu} \rangle$ in these impact parameter
bins are in relatively good agreement with those determined by BRAHMS.
The results in the two lowest
$\eta$ bins are reported for $(h^+ + h^-)/2$ while the $\eta = 2.2$ 
and 3.2 bins
are reported for $h^-$ only where $h^+$ and $h^-$ stand for the positively and
negatively charged hadrons respectively.

In this paper, we calculate $R_{\rm dAu}(p_T)$ in the BRAHMS $\eta$ bins using
two parameterizations of nuclear shadowing.  We also calculate the
central-to-peripheral ratios, $R_{\rm CP}(p_T)$, with two parameterizations of
the spatial dependence of shadowing.  The calculated 
ratios are compared to the BRAHMS data \cite{brahmsdau}.
To better illustrate the effects of shadowing alone, we do
not include the Cronin effect, $p_T$ broadening
\cite{Gyulassy:2001nm,Vitev:2002vr}, in our calculations.

We make a leading order (LO) calculation of minijet production to 
obtain the yield of high-$p_T$ partons \cite{usprc}.  The $p_T$ distribution
of Ref.~\cite{Field} is modified to include the nuclear parton distribution
functions, 
\begin{eqnarray}
\frac{d\sigma_{{\rm dAu} \rightarrow hX}}{d^2b dp_T} & = & 
2p_T \sum_{i,j = q, \overline q, g}
\int_{\theta_{\rm min}}^{\theta_{\rm max}} 
\frac{d\theta_{\rm cm}}{\sin
\theta_{\rm cm}} \int dx_1 \int dx_2 \int d^2s \, \int dz \, \int dz' \,  
\nonumber \\
&  & \mbox{} \times F_{i/{\rm d}}(x_1,Q^2,\vec s,z) 
F_{j/{\rm Au}}(x_2,Q^2,|\vec b - \vec s|, z')
\frac{D_{h/k}(z_c,Q^2)}{z_c} \frac{d\hat \sigma_{ij \rightarrow k}}{d 
\hat t} \, \,  
\end{eqnarray}
where $x_1$ and $x_2$ are the parton momentum fractions in the deuterium and
gold nuclei respectively, $Q$ is the momentum scale of the hard interaction and
$z_c$ is the fraction of the parton momentum transferred to the final-state
hadron. 
The integrals over center-of-mass scattering angle, $\theta_{\rm min} \leq 
\theta_{\rm cm} \leq \theta_{\rm max}$, 
correspond to the BRAHMS angular regions, given previously.
The $2 \rightarrow 2$ minijet cross sections, 
$d\hat \sigma_{ij \rightarrow k}/d \hat t$, are given in Ref.~\cite{owens}.
Even though the next-to-leading order corrections may affect the shape of the
$p_T$ distributions, the higher-order
corrections should largely cancel out in $R_{\rm dAu}$, 
as is the case for $J/\psi$
\cite{psiprl} and Drell-Yan \cite{usprc} production.  

The parton densities in the gold nucleus, 
$F_{j/{\rm Au}}(x,Q^2,\vec{b},z)$, can be
factorized into $x$ and $Q^2$ independent nuclear density
distributions, position and nuclear-number independent nucleon parton
densities, and a shadowing function, 
$S^j_{{\rm P},{\rm S}}(A,x,Q^2,\vec{b},z)$,
that describes the modification of the nuclear parton distributions in
position and momentum space.  The first subscript on the shadowing function, 
P, refers to the shadowing parameterization
while the second,  S, to the spatial dependence.
Most available shadowing parameterizations ignore
effects in deuterium so that $F_{i/{\rm d}}$ depends only on the deuterium 
density distribution and the nucleon
parton densities.  We account for the proton and neutron
numbers of both nuclei.
Then \cite{psiprl}
\begin{eqnarray}
F_{i/{\rm d}}(x,Q^2,\vec{s},z) & = & \rho_{\rm d}(\vec s,z) f_{i/N}(x,Q^2) 
\label{fadeut} \\
F_{j/{\rm Au}}(x,Q^2,|\vec b - \vec s|,z') & = & \rho_{\rm Au}(|\vec b -
\vec s|,z') 
S^j_{{\rm P},{\rm S}}({\rm Au},x,Q^2,|\vec b - \vec{s}|,z') 
f_{j/N}(x,Q^2) \label{nucglu} \, \, 
\end{eqnarray}
where $f_{i/N}(x,Q^2)$ is the nucleon parton density. 
In the absence of nuclear
modifications, $S^j_{{\rm P},{\rm S}} \equiv 1$.  The nucleon density
distribution of the 
gold nucleus is assumed to be a Woods-Saxon with $R_{\rm Au} =
6.38$ fm \cite{DeJager:1974dg}.   We use the Hulth\'en
wave function \cite{hulthen} to calculate the deuteron density
distribution.  The densities are normalized so that $\int d^2s dz
\rho_A(\vec s,z) = A$.  We use
the MRST LO parton distributions \cite{Martin:1998np} for isolated nucleons and
take $Q^2 = p_T^2$.  

We have chosen two parameterizations of nuclear shadowing
which cover extremes of gluon shadowing at low $x$.  The Eskola
{\it et al.} parameterization, EKS98, is based on the GRV LO
\cite{Gluck:1991ng} parton densities.  At the minimum scale, $Q_0^2$, valence 
quark shadowing is identical for
$u$ and $d$ quarks.  Likewise, $\overline u$, $\overline d$ and $\overline s$
shadowing is identical at $Q_0^2$.  Even though the light quark
shadowing ratios are not constrained to be equal at higher scales, the 
differences between them are small.  Shadowing of the heavier flavor
sea, $\overline s$ and higher, is calculated separately at $Q_0^2$.  The
shadowing ratios for each parton type are evolved to LO for $2.25 < Q^2 <
10^4$ GeV and are valid for 
$x \geq 10^{-6}$ \cite{Eskola:1998iy,Eskola:1998df}.
Interpolation in nuclear mass number allows results to be obtained for
any input $A$.  The parameterizations by Frankfurt, Guzey and
Strikman combine Gribov theory with hard
diffraction \cite{Frankfurt:2003zd}.  They are based on the CTEQ5M 
\cite{Lai:1999wy} parton
densities and evolve each parton species separately to NLO for $4 < Q^2
< 10^4$ GeV.  Although the $x$ range is $10^{-5} < x < 0.95$, the
sea quark and gluon ratios are unity for $x > 0.2$.  The EKS98 valence
quark shadowing ratios are used as input since Gribov theory does not
predict valence shadowing.  The parameterizations are available for four
different values of $A$: 16, 40, 110 and 206.  We use $A = 206$
for the gold nucleus with the parameterization that gives
the strongest gluon shadowing for the largest contrast to EKS98, 
denoted FGS1 here.

We now turn to the spatial dependence of the shadowing.  Since some qualitative
spatial dependence has been observed \cite{E745} but the exact behavior is
unknown, we have tried two different parameterizations for inhomogeneous
shadowing in d+Au collisions \cite{usprc,us,firstprl,spenpsi}.  The
first, $S^j_{{\rm P}, {\rm WS}}$, assumes that shadowing is proportional
to the local density, $\rho_A(r)$,
\begin{eqnarray}
S^j_{{\rm P},{\rm WS}}(A,x,Q^2,\vec s,z) & = & 1 + N_{\rm WS}
[S^j_{\rm P}(A,x,Q^2) - 1] \frac{\rho_A(r)}{\rho_A(0)} \label{wsparam} \, \, ,
\end{eqnarray}
where $r = \sqrt{s^2 + z^2}$, 
$\rho_A(0)$ is the central density and 
$N_{\rm WS}$ is chosen so
that $(1/A) \int d^2s dz \rho_A(\vec s,z) S^j_{{\rm P},{\rm WS}} = S^j_{\rm
P}$. When $r \gg R_A$, the nucleons behave as free particles while, at
the center of the nucleus, the modifications are larger than $S^i_{\rm P}$.

If, instead, shadowing stems from multiple interactions of
the incident parton \cite{ayala}, parton-parton interactions are spread
longitudinally over the coherence length, $l_c=1/2m_Nx$, where $m_N$
is the nucleon mass \cite{ina}.  For $x<0.016$, $l_c >R_A$ for any $A$ and
the incident parton interacts coherently with all the target partons
in its path so that
\begin{eqnarray}
S^j_{{\rm P},\rho}(A,x,Q^2,\vec{s},z) = 1 + N_\rho [S^j_{\rm P}(A,x,Q^2) - 1]
\frac{\int dz \rho_A(\vec{s},z)}{\int dz \rho_A(0,z)} \label{rhoparam} \, \, . 
\end{eqnarray}
The integral
over $z$ includes the material traversed by the incident nucleon.  The
normalization requires $(1/A) \int d^2s dz \rho_A(\vec s, z) S^j_{{\rm
P},\rho} = S^j_{\rm P}$ with $N_\rho > N_{\rm WS}$.
At large $x$, $l_c\ll R_A$ and shadowing is
proportional to the local density, Eq.~(\ref{wsparam}).  

While there are three homogeneous FGS parameterizations, only two inhomogeneous
parameterizations are provided.  No spatial dependence is given for
FGS1, the case with the strongest gluon shadowing.  We have checked the 
available dependencies against those calculated using $S^j_{\rm FGS1,WS}$ and 
$S^j_{{\rm FGS1},\rho}$ and found that, at 
similar values of
the homogeneous shadowing ratios, $S^j_{{\rm FGS1},\rho}$ is quite 
compatible with the available FGS inhomogeneous parameterizations.  
Therefore, to characterize the spatial dependence
of FGS1, we use $S_{{\rm FGS1},\rho}^j$.

The fragmentation functions, $D_{h/k}(z_c,Q^2)$, describe the
production of hadron $h$ from parton $k$ with $z_c = p_h/p_k$. 
The produced partons are fragmented into charged pions, kaons and
protons using the LO KKP fragmentation functions \cite{Kniehl:2000fe},
fit to $e^+ e^-$ data.  The final-state hadrons are assumed to be produced
pairwise so that $\pi \equiv (\pi^+ + \pi^-)/2$, $K \equiv (K^+ +
K^-)/2$, and $p \equiv (p + \overline p)/2$.  The equality of $p$ and
$\overline p$ production obviously does not describe low energy hadroproduction
well.  At higher energies, however, the approximation that $p = \overline p$ 
may be more reasonable.  The produced hadrons follow the parent parton
direction.  The minimum $Q^2$ in the KKP fragmentation functions is 
$Q_{{\rm Fr}_0}^2 = 2$ GeV$^2$, 
similar to but somewhat lower than the minimum $Q^2$
of the shadowing parameterizations.
Thus the minimum $p_T$ of our calculations is $\sqrt{2}$ GeV.
We assume the same scale in the parton densities 
and the fragmentation functions, $Q^2 = Q_{\rm Fr}^2 = 
p_T^2$.  A larger scale, $p_T^2/z_c^2$, is sometimes used in the parton
densities but where $z_c$ is large, as is the case here,
changing the scale does not significantly alter the calculated ratios.

The largest contribution to the total final-state charged particle production
is from
the charged pions, followed by the kaons.  The proton contribution is the
smallest even though, in d+Au collisions at RHIC, $(p + \overline p)/h \approx
0.24 \pm 0.02$ where $h = h^+ + h^-$ for $2<p_T<3$ GeV, independent of
centrality \cite{Sakaguchi:2002bm}.  The d+Au result is similar to that from
$pp$, $0.21 \pm 0.01$ \cite{Sakaguchi:2002bm}.  
The discrepancy between the RHIC d+Au and $pp$ results
and the extrapolation from $e^+ e^-$ is due to the poor knowledge of the 
fragmentation functions at large $z_c$.

We have calculated the $p_T$ distributions for final-state charged pions, kaons
and protons/antiprotons separately as well as the sum of all charged particles.
For each final-state hadron, we determine the fraction of the total from
produced quarks, antiquarks and gluons.  In the central $\eta$ bin for pion
production, gluons are produced 
almost equally in the $gg \rightarrow gg$ and $qg
\rightarrow qg$ channels.  The $qg$ channel is somewhat larger for $p_T
> 5$ GeV.  There is a negligible contribution from $q \overline q
\rightarrow gg$.  Pion production by quarks and antiquarks proceeds mainly
through the $qg \rightarrow qg$ channel for quarks
and $\overline q g \rightarrow \overline qg$ for antiquarks.  The next
largest contribution to pion production by quarks are the 
$q q' \rightarrow q q'$ and $qq \rightarrow q q$
channels which are of very similar strength,
followed by $q \overline q \rightarrow
q \overline q$ and $gg\rightarrow q \overline q$ with a negligible contribution
from $q \overline q \rightarrow q' \overline q'$.  The contributions to
antiquark production after $\overline q g \rightarrow \overline q g$ are
$\overline q q' \rightarrow \overline q q'$, $\overline q q 
\rightarrow \overline q  q$ and $gg \rightarrow q \overline q$, followed by
smaller contributions from $\overline q \overline q \rightarrow \overline q
\overline q$ and $q \overline q \rightarrow q' \overline q'$.
Similar results are found for kaon and
proton production.  However, the proton distributions fall off more steeply
with $p_T$.

The relative contributions from the production channels remain similar
as rapidity increases.  The most important change is in gluon production where 
the $qg \rightarrow qg$ channel grows more dominant, finally becoming larger
than the $gg \rightarrow gg$ channel for all $p_T$ for $\eta = 3.2$.  Indeed,
at the most forward rapidity, the $q \overline q \rightarrow gg$ channel 
becomes comparable to the $qg$ channel at $p_T \sim 8$ GeV.
This may seem counterintuitive since the ion $x_2$
value decreases as $\eta$ grows, 
increasing the gluon density.  However, the deuteron
$x_1$ value increases more rapidly and, at large $\eta$, we are in a region
where the deuteron gluon density is dropping steeply while the quark density,
particularly that of the valence quarks, is still significant.  Thus the $qg$
channel is more important than the $gg$ channel at large $\eta$, particularly 
when $p_T$ and $x_1$ are large.  Also, at high $\eta$ and $p_T$, $p_T > 7.5$ 
GeV, antiquarks are predominantly produced by valence quark induced processes 
since these are large at high $x$.

Because we begin to approach the edge of phase space with increasing $\eta$,
the $p_T$ distributions steepen, especially for antiquark and gluon production.
Quark production, which includes the valence contribution, dominant at high
$p_T$ and $\eta$, remains harder overall.  Thus quark production will come
to dominate all final-state hadron production.  This effect,
increasingly important at high $p_T$ and $\eta$, is reflected in the relative
contributions to pion, kaon and proton production by quarks, antiquarks and
gluons.  At $|\eta| \leq 0.2$, pion production is dominated by
produced gluons up to $p_T > 9$ GeV where pion production by quarks becomes
larger.  Gluon production of kaons is rather small, similar to the quark
contribution at $p_T \sim Q_{{\rm Fr}_0}$ but dropping below the antiquark
contribution at $p_T \sim 3.5$ GeV.  Quark production is most important for
protons at $p_T > 3.5$ GeV.  As $\eta$ increases, quark production of 
final-state hadrons becomes increasingly dominant.  Already 
at $\eta = 1$, more 
than half of all kaons and protons are produced by quarks for $p_T > 2.5$ GeV.
At higher $\eta$, antiquarks and gluons make negligible contributions to 
low $p_T$ kaon and proton production at $\eta = 2.2$ and 3.2.  Quarks also
dominate pion production for $p_T > 6.5$, 3 and 1.5 GeV with $\eta = 1$, 2.2 
and 3.2 respectively.

We have calculated the average Au ion
momentum fraction, $\langle x_2 \rangle$, and the average deuteron momentum
fraction, $\langle x_1 \rangle$, for $\sqrt{2} \leq p_T \leq 12$ GeV.  
However, the
largest accessible $p_T$ decreases to 9.5 GeV at $\eta = 3.2$ due to phase
space.  The results are shown in Table~\ref{avetab}.  
Since these are average $x_2$ values, the actual $x_2$ for each event 
can be smaller or larger than these averages.  The minimum and maximum $\langle
x_2 \rangle$ correspond to the lowest and highest $p_T$ values respectively.
Both the minimum and maximum values decrease as $\eta$ increases so that the
minimum $\langle x_2 \rangle$ is reached at $\eta = 3.2$.  However, the maximum
$\langle x_2 \rangle$ increases relative to more central $\eta$ values due to
the reduction of phase space at high $p_T$.  Note that as $\theta \rightarrow
0$, $\langle x_2 \rangle \rightarrow 1$.  The averages are not very sensitive
to changes in the parton densities or the choice of factorization, 
renormalization or fragmentation scales.

The total hadron yield closely follows that of the pions.  There
is little variation of $\langle x_2 \rangle$ between hadron species although
the proton averages are generally somewhat smaller than those of the mesons.  
A small difference between the partonic contributions to hadron production
can be attributed to the behavior of the parton distribution functions
in the various production channels.  A set of LO parton densities derived 
including 
the GLRMQ recombination terms at low $x$ found deviations from normal DGLAP 
evolution at $x < 10^{-3}$ for the proton \cite{EHKQS}.  
Thus one may question whether saturation effects
can be at work here when $\langle x_2 \rangle = 0.035$
at $\eta = 3.2$.  

While the average Au momentum fraction
is decreasing with centrality, the average deuteron momentum fraction, $x_1$, 
is increasing.   Note that the maximum $\langle x_1 \rangle$ at $\eta = 3.2$
approaches 1, indicative of the edge of phase space.
The average $z_c$ in the fragmentation functions is large, $\approx 0.5$ at
midrapidity, and 
increasing with $\eta$ and $p_T$.  The fragmentation
functions are best determined for smaller $z_c$ so that the high $z_c$
fragmentation functions are unreliable, especially for baryon production.
Modeling of high $p_T$ and high $\eta$ hadron production thus contains large
theoretical uncertainties due to the fragmentation functions.  
There is more variation in $\langle z_c \rangle$ due to parton
type than in $\langle x_2 \rangle$.  The proton $\langle z_c \rangle$ tends
to be somewhat smaller than for pions or kaons.  The tabulated values are
for total charged hadrons.

\begin{figure}
\setlength{\epsfxsize=0.9\textwidth}
\setlength{\epsfysize=0.6\textheight}
\centerline{\epsffile{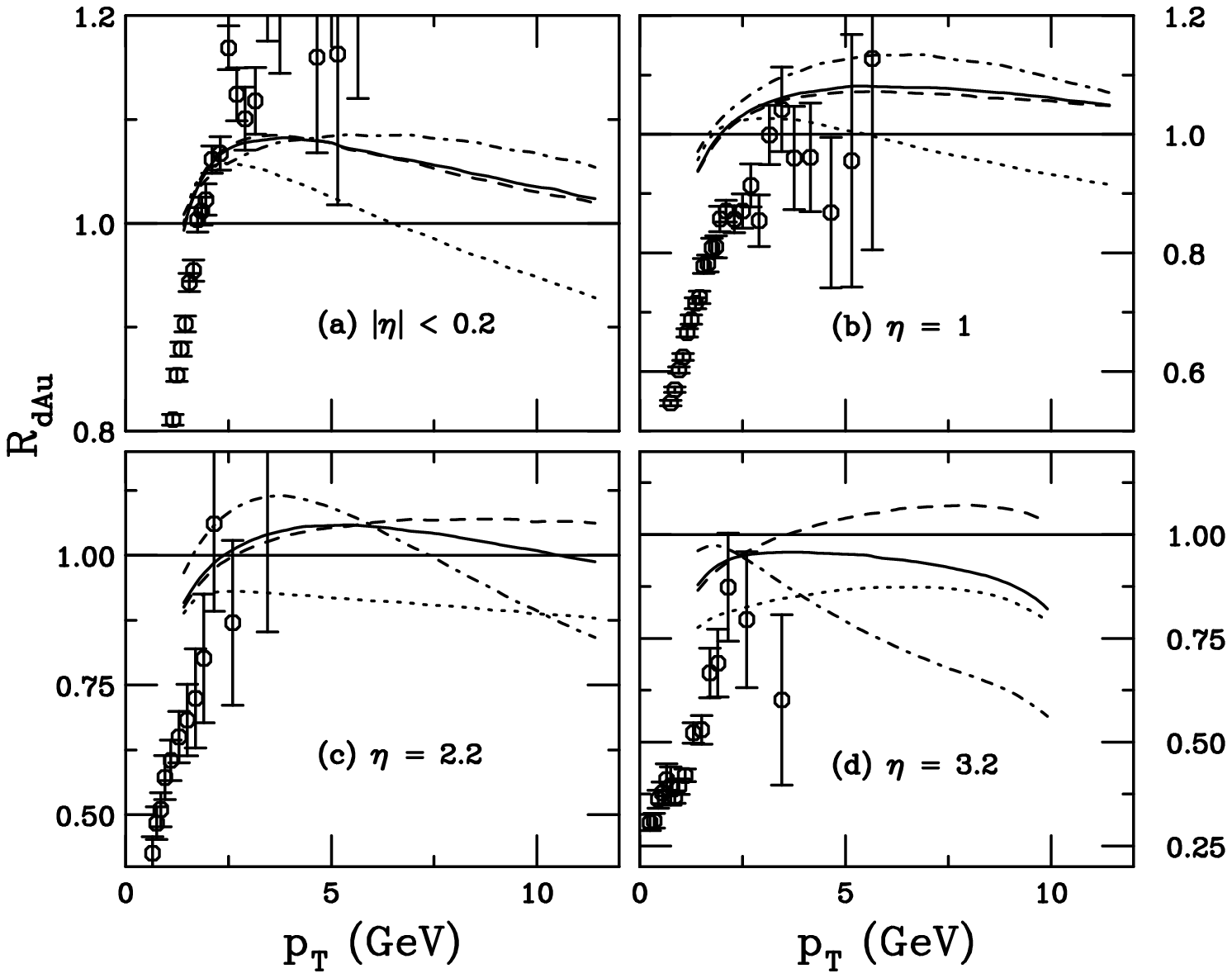}}
\caption{$R_{\rm dAu}$ for charged pions (dashed) and kaons (dot-dashed) as
well as protons and antiprotons (dotted) and the sum over all charged 
hadrons (solid) for deuteron-gold collisions at $\sqrt{S_{NN}} = 200$ GeV 
as a function of $p_T$.  The results for homogeneous shadowing with
the EKS98 parameterization are compared to the minimum bias BRAHMS data 
\protect\cite{brahmsdau} in
the following $\eta$ bins: (a) $|\eta| \leq 0.2$;
(b) $\eta = 1$; (c) $\eta = 2.2$ and (d) $\eta = 3.2$.}
\label{rdaueks98}
\end{figure}

\begin{figure}
\setlength{\epsfxsize=0.9\textwidth}
\setlength{\epsfysize=0.6\textheight}
\centerline{\epsffile{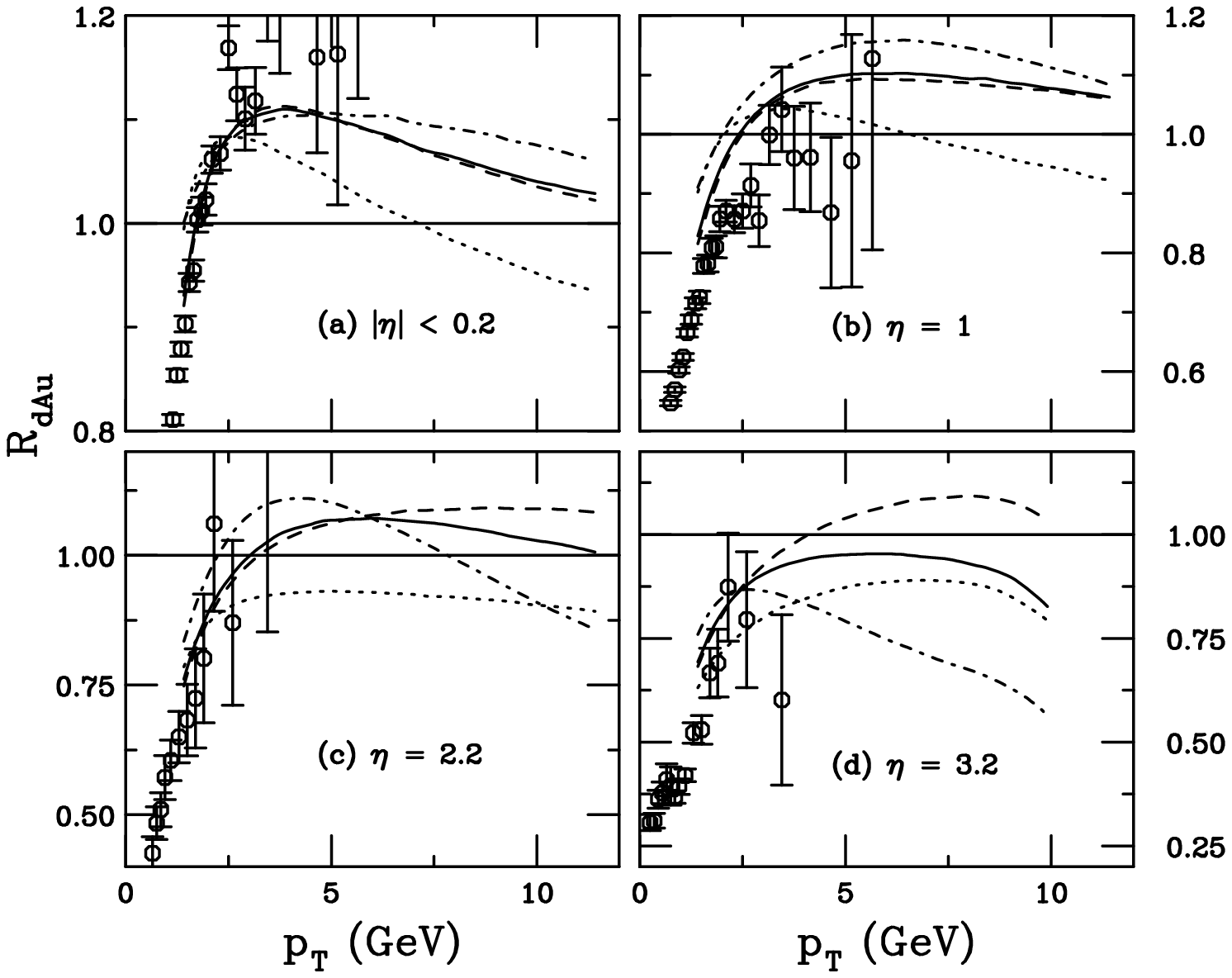}}
\caption{$R_{\rm dAu}$ for charged pions (dashed) and kaons (dot-dashed) as
well as protons and antiprotons (dotted) and the sum over all charged 
hadrons (solid) for deuteron-gold collisions at $\sqrt{S_{NN}} = 200$ GeV 
as a function of $p_T$.  The results for homogeneous shadowing with
the FGS1 parameterization are compared to the minimum bias BRAHMS data 
\protect\cite{brahmsdau} in
the following $\eta$ bins: (a) $|\eta| \leq 0.2$;
(b) $\eta = 1$; (c) $\eta = 2.2$ and (d) $\eta = 3.2$.}
\label{rdaufgs}
\end{figure}

We now compare the ratios, $R_{\rm dAu}$, calculated for the two homogeneous
shadowing parameterizations, to the BRAHMS data \cite{brahmsdau}.  
The EKS98 results are shown for each $\eta$
interval in Fig.~\ref{rdaueks98} while those employing 
the FGS1 parameterization
are shown in Fig.~\ref{rdaufgs}.  We show the results for charged pions
(dashed), charged kaons (dot-dashed) and protons/antiprotons (dotted)
separately.  The solid curves give the total charged hadron result.
At midrapidity, where $\langle x_2 \rangle$ is
relatively large, the two parameterizations give rather similar results.
As pointed out in Ref.~\cite{kvraa}, the difference between the kaon and proton
ratios is due to isospin effects.  As long as pion production is dominated 
by gluons, it is essentially 
independent of isospin.  The ratio is greater than 
unity but smaller than the BRAHMS result at
midrapidity.  Including $p_T$ broadening would increase the $|\eta| \leq 0.2$
ratio.  

At $\eta = 1$ and low $p_T$, the ratio is less than unity for both
parameterizations but the stronger gluon shadowing in the FGS1 parameterization
reduces $R_{\rm dAu}$ to $\sim 0.8$ for $p_T = \sqrt{2}$ GeV relative to $\sim
0.9$ for EKS98.  At $p_T \sim 2.5$ GeV, $R_{\rm dAu}$ rises above unity again.
At higher rapidities, $R_{\rm dAu}$ decreases at low $p_T$ but does not rise
as far above unity at higher $p_T$ until, at $\eta = 3.2$, the
total charged hadron ratio is less than unity for all $p_T$.  The EKS98
parameterization tends to underestimate the data for all but the most central 
rapidities, see Fig.~\ref{rdaueks98}.  The FGS1 parameterization, 
on the other hand, agrees rather
well with the central $\eta$ data, Fig.~\ref{rdaufgs}(a), and lies within the
errors of the most peripheral bins for $p_T > \sqrt{2}$ GeV, 
Fig.~\ref{rdaufgs}(c) and (d).  However, the total
charged hadron data at $\eta = 1$
are somewhat underestimated by the FGS1 parameterization,
Fig.~\ref{rdaufgs}(b).  

In the most central bins, the ratio for the total charged hadrons closely
follows $R_{\rm dAu}$ for the pions.  At higher $\eta$, the kaon contribution
becomes more important, causing the total to be closer to the average of the
pion and kaon results.  The proton contribution, on the other hand, remains
small, even at $\eta = 3.2$, while one may expect that, in reality, proton
production would be more important at large rapidity as the 
fragmentation region is 
approached. However, this effect cannot be accounted for by standard
parameterizations of the fragmentation functions.
Since BRAHMS measures the negative charged hadron 
distribution, $h^-$, at $\eta = 2.2$ and 3.2, only the antiprotons can
contribute to $R_{\rm dAu}$.  

We point out that because $x$ is not very small for $p_T$ larger than a few
GeV and the shadowing ratios are also not far from unity in this region,
$R_{\rm dAu}$, the ratios of d+Au relative to $pp$, are driven by isospin
rather than shadowing.  This is obvious from the very similar behavior of the
EKS98 and FGS1 ratios seen in Figs.~\ref{rdaueks98} and \ref{rdaufgs} for
$p_T > 5$ GeV.

\begin{figure}
\setlength{\epsfxsize=0.9\textwidth}
\setlength{\epsfysize=0.6\textheight}
\centerline{\epsffile{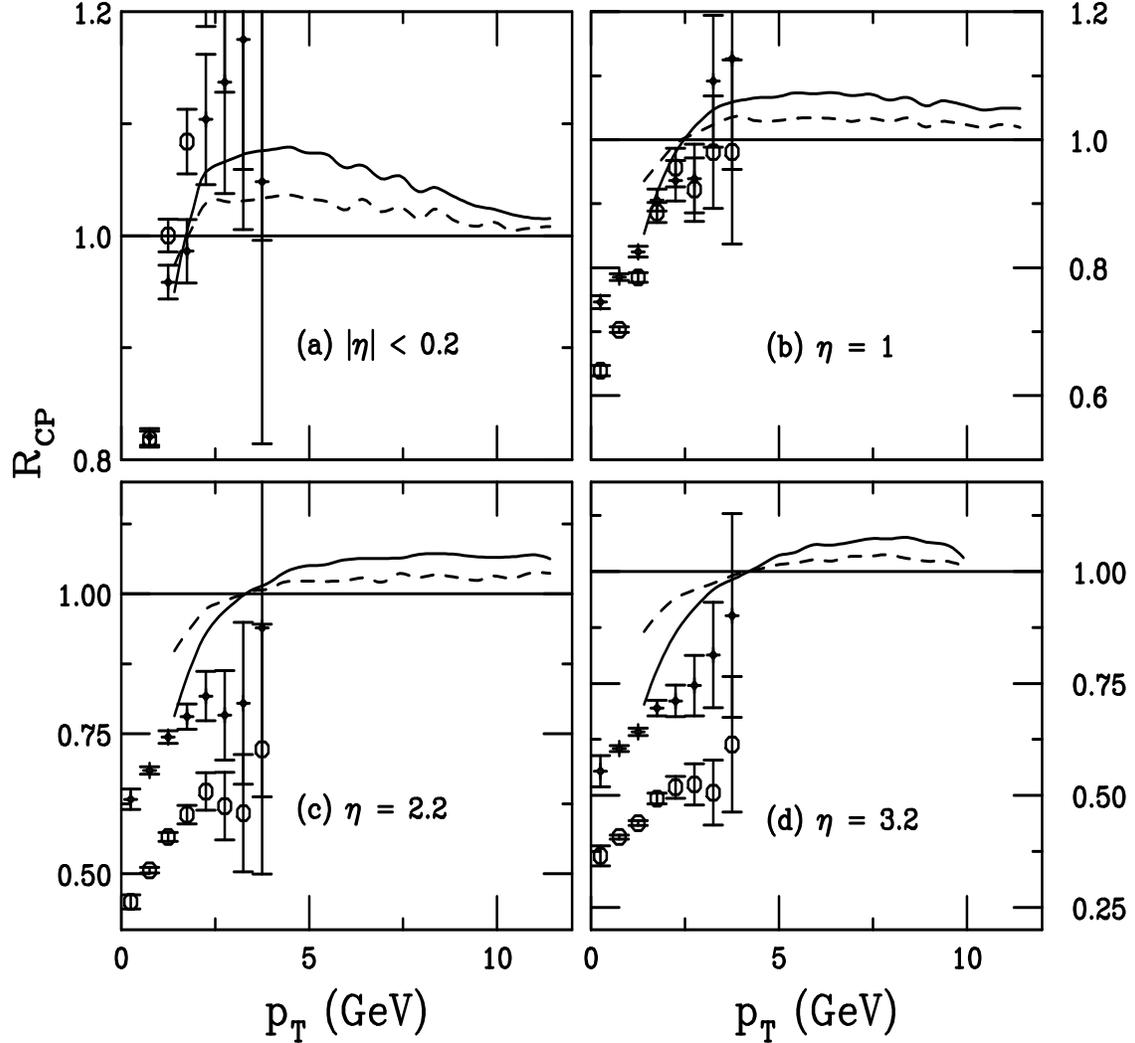}}
\caption{$R_{\rm CP}$ for charged hadrons in deuteron-gold collisions at 
$\sqrt{S_{NN}} = 200$ GeV 
as a function of $p_T$.  The results for $S_{{\rm FGS1},\rho}$ 
are compared to the BRAHMS data \protect\cite{brahmsdau} in
the following $\eta$ bins: (a) $|\eta| \leq 0.2$;
(b) $\eta = 1$; (c) $\eta = 2.2$ and (d) $\eta = 3.2$.  The calculated ratios
of the most central and semi-central to peripheral collisions are shown in the
solid and dashed curves, respectively.  The BRAHMS data are given by the open
circles (most central) and diamonds (semi-central).}
\label{rcpsrho}
\end{figure}

Figure \ref{rcpsrho} illustrates the centrality dependence using 
$S^j_{{\rm FGS1}, \rho}$. We compare the central-to-peripheral ratio, $R_{\rm
CP}$, which should be less sensitive to isospin than $R_{\rm dAu}$, to
the BRAHMS data.  The solid curves show the ratio
of the central, $(0-20)$\%, to peripheral, $(60-80)$\%, bins 
for each $\eta$ region while the dashed
curves show the semi-central, $(30-50)$\%, to peripheral ratios.
Our calculations assume exact impact parameter cuts while, experimentally,
impact parameter is poorly measured on an event-by-event basis in d+Au
collisions.  We note that both of the inhomogeneous EKS98 results 
are much weaker than those in Fig.~\ref{rcpsrho} and are not shown here.

In central collisions, with small impact parameter, inhomogeneous
shadowing is stronger than the homogeneous result.  The larger the homogeneous
shadowing effect, the larger the difference between $S^j_{\rm P}$ and $S^j_{\rm
P,S}$.  Thus $R_{\rm CP}$ is a stronger function of impact parameter for the
FGS1 parameterization since it has larger homogeneous shadowing at small $x$.
The ratios with $S^j_{\rm EKS,S}$ underestimate the centrality dependence 
considerably and are not shown.  Note
that $R_{\rm CP}$ approaches unity at large $p_T$ since the difference
between $S^j_{\rm P}$ and $S^j_{\rm P,S}$ decreases as $x$ increases and the
shadowing effect becomes small.

Since $S^j_{{\rm FGS1},\rho}(b)$ is a rather smooth function of impact 
parameter, the $b$ dependence of $R_{\rm CP}$ is not very strong.  
The fluctuations in $R_{\rm CP}$ for $S^j_{{\rm FGS1},\rho}$, especially
notable at central rapidity, are due to the discrete steps of $T_{\rm Au}(r)$
in the integration over the spatial coordinates.  These fluctuations are absent
for $S^j_{\rm P,WS}$ since $\rho_{\rm Au}(r)$ is a smooth function.

The agreement with the data is reasonable at central
$\eta$, see Fig.~\ref{rcpsrho}(a) and (b).  The trends of the impact parameter
dependence are similar to the data at low $p_T$.  
The semi-central-to-peripheral
ratio is similar to the central-to-peripheral ratio in the most central 
rapidity bin while in the more peripheral bins, the central-to-peripheral
ratio has a stronger $p_T$ dependence.  However, the increase in $\langle
x_2 \rangle$ with $p_T$ results in the strong growth of $R_{\rm CP}$ with $p_T$
at forward $\eta$.  The resulting curvature of the calculated ratio is faster
than the data.  The magnitude of $R_{\rm CP}$
at low $p_T$ is also underestimated.  Since the position
dependence of inhomogeneous shadowing is not well understood, the poorer
agreement with the centrality-dependent data in Fig.~\ref{rcpsrho} 
compared to the minimum bias results in Figs.~\ref{rdaueks98}
and \ref{rdaufgs} is not surprising.
These data could be used to tune the position dependence of shadowing.

In summary, we find that the suppression factor, $R_{\rm dAu}$, calculated with
leading-twist shadowing, especially employing the FGS1 parameterization,
agrees moderately well with the BRAHMS data.
These calculations imply that saturation effects may not play a dominant role
in the forward region at RHIC, as suggested in other recent work 
\cite{Jamal,Dima}.
Our calculations of $R_{\rm CP}$ show a stronger $p_T$ dependence than that
suggested by BRAHMS, likely due to insufficient
data on the impact parameter dependence of nuclear shadowing.

We thank K.J. Eskola and V. Guzey
for providing the shadowing routines and J. Gonzalez, S. R. Klein, 
M. Murray, M. Strikman and W. Vogelsang for
discussions.  This work 
was supported in part by the Division of Nuclear Physics of the Office
of High Energy and Nuclear Physics of the U. S. Department of Energy
under Contract Number DE-AC03-76SF0098.

\newpage
\begin{table}[htb]
\begin{tabular}{ccc|cc|cc}
$\langle \eta \rangle$ & \multicolumn{2}{c|}{$\langle x_2 \rangle$} 
& \multicolumn{2}{c|}{$\langle x_1 \rangle$} & \multicolumn{2}{c}{$\langle 
z_c \rangle$} \\
 & min & max & min & max & min & max \\ \hline
$\approx 0$ & 0.07 & 0.22 & 0.07 & 0.22 & 0.52 & 0.64 \\
1 & 0.055 & 0.18 & 0.1 & 0.33 & 0.54 & 0.68 \\
2.2 & 0.042 & 0.14 & 0.17 & 0.62 & 0.55 & 0.81 \\
3.2 & 0.035 & 0.23 & 0.32 & 0.95 & 0.64 & 0.96 \\ 
\end{tabular}
\caption[]{The average values of the Au and d momentum fractions, $\langle x_2
\rangle$ and $\langle x_1 \rangle$ respectively, as well as the average 
fraction of the final-state parton momentum transferred to the hadron, $\langle
z_c \rangle$, in the four
BRAHMS pseudorapidity intervals.  The minimum values correspond to $p_T \approx
\sqrt{2}$ GeV while the maximum corresponds to $p_T = 12$ GeV for the first
three $\eta$ bins and 10 GeV for the most forward $\eta$ bin.}
\label{avetab}
\end{table}

\newpage
\newpage

\newpage

\end{document}